# Generic Galilean invariant exchange correlation functionals with quantum memory

Yair Kurzweil and Roi Baer♦

*Department of Physical Chemistry and the Lise Meitner Center for Quantum Chemistry, the Hebrew University of Jerusalem, Jerusalem 91904 Israel.*

Today, most application of time-dependent density functional theory (TDDFT) use adiabatic exchange-correlation (XC) potentials that do not take into account non-local temporal effects. Incorporating such "memory" terms into XC potentials is complicated by the constraint that the derived force and torque densities must integrate to zero at every instance. This requirement can be met by deriving the potentials from an XC action that is Galilean invariant (GI). We develop a class of simple but flexible forms for an action that respect these constraints. The basic idea is to formulate the action in terms of the Eularian-Lagrangian transformation (ELT) metric tensor, which is itself GI. The general form of the XC potentials in this class is then derived and the linear response limit is derived as well.

## I. INTRODUCTION

Time dependent density functional theory (TDDFT)[1] is routinely used in many calculations of electronic processes in molecular systems. Almost all applications use "adiabatic" potentials describing an immediate response of the Kohn-Sham potential to the temporal variations of the electron density. The shortcomings of these potentials were studied by several authors[2-5]. Some of the problems are associated with self interaction, an ailment inherited from ground-state density functional theory[6]. Other deficiencies are known or suspected to be associated with the adiabatic assumption. The first attempt to include non-adiabatic effects[7] was based on a simple form of the exchange-correlation (XC) potential in the linear response limit. Studying an exactly solvable system, this form was shown to lead to spurious time-dependent evolution[8]. The failure was traced back to violation of a general rule: the XC force density, derived from the potential, should integrate to zero[9]. Convincing arguments were then presented[10], demonstrating that non-adiabatic effects cannot be easily described within TDDFT and instead a *current density* based theory must be used. Vignale and Kohn[10] gave an expression for the XC potentials applicable for linear response and long wave lengths.

That the total XC force is zero is a valid fact not only in TDDFT but also in TDCDFT. It stems from the basic requirement that the total force on the non-interacting particles must be equal to the total force on the interacting particles. This is so otherwise a different total acceleration results and the two densities or current densities will be at variance. In the interacting system the total (Ehrenfest) force can only result from an external potential: because of Newton's third law the electrons cannot exert a net force upon themselves. In TDCDFT the total force equals the sum of the external force, the Hartree force and the XC force. Since the Hartree force integrates to zero (Newton's third law again) the total XC force does so as well. A similar general argument can be applied to the total torque, showing that the net XC torque must be zero. These requirements then have to be imposed on the approximate XC potentials[9].

The question we deal with in this paper is the how to construct approximations to the XC potentials that manifestly obey the zero XC force and torque condition. For this purpose, we develop the concept of a XC "action", obtaining the potentials as functional derivatives of such an action. This is similar to the concepts of DFT, where the XC potential is functionally derived from an energy functional. Our action is a functional $S[\mathbf{u}]$ of the electron fluid velocity field $\mathbf{u}(\mathbf{r},t) = \mathbf{j}(\mathbf{r},t)/n(\mathbf{r},t)$ where $n(\mathbf{r},t)$ and $\mathbf{j}(\mathbf{r},t)$ are the particle and current densities. It also depends on the initial electron density $n_0(\mathbf{r})$. The reader may wonder why our action depends only on $\mathbf{u}$ and not on the current density $\mathbf{j}$, which is the natural variable in TDCDFT[10]. We have found that working with the velocity field is easier to construct GI functionals. Furthermore, this is no violation of the Runge-Gross theorem since knowledge of $n_0(\mathbf{r})$ and $\mathbf{u}(\mathbf{r},t)$ is sufficient to uniquely determine $\mathbf{j}(\mathbf{r},t)$ (and $n(\mathbf{r},t)$) for $t>0$, by solving the following pair of equations (the second of which is the continuity equation, which must hold in TDCDFT):

$$\begin{aligned}\mathbf{j}(\mathbf{r},t) &= n(\mathbf{r},t)\mathbf{u}(\mathbf{r},t) \\ \dot{n}(\mathbf{r},t) + \nabla\cdot\mathbf{j}(\mathbf{r},t) &= 0, \quad n(\mathbf{r},0) = n_0(\mathbf{r})\end{aligned} \quad (1.1)$$

The critical point is, that by ensuring that the action is "Galilean Invariant", the derived potentials are assured to obey the zero force and torque conditions[9, 11]. But what do we mean by a "Galilean Invariant" (GI) action? It means that any two observers of the same physical system (each using his own coordinate system of reference) will report the same value for the action. We define GI in detail in section II. Galilean frames can be translationally or rotationally accelerating. Invariance in the first case is called translational invariance (TI) and in the second case, rotational invariance (RI). We discuss this in more detail in section II. Kurzweil and Baer[11] have recently developed a a TI XC action and derived potentials that obey the zero force condition. Their XC action was however not RI and so did not enforce the zero torque condition. It is the purpose of this paper to further develop the theory along similar lines, to achieve zero XC torque as well. We limit our discussion to as simple a theory as possible, by considering as building blocks only low order derivatives of basic quantities. Simple generic forms of the XC action are developed in section III. The detailed derivation of XC potentials from them is then developed

♦ Corresponding author: FAX: +972-2-6513742, roi.baer@ huji.ac.il

in section IV. The result of that section is a reasonably general form of XC potential that conforms to the zero torque and force conditions. In section V we connect the general theory to known results for linear response of the homogeneous electron gas. Arriving at a plausible form of the action functional that is compatible with the HEG longitudinal and transverse response functions.

## II.  GALILEAN INVARIANT ACTION

As noted above, Galilean invariance of the action means that observers in different Galilean frames report the same value for the XC action. To further explain this point, let us consider two types of relative motion: translational and rotational. One observer, using "unprimed" coordinates, denotes the current density as $\mathbf{j}(\mathbf{R},t)$ and particle density as $n(\mathbf{R},t)$. A second observer is using primed coordinates and its coordinate origin is accelerating with respect to the first observer's origin. A given point in space designated as $\mathbf{R}$ by the first observer is designated by

$$\mathbf{R}' = \mathbf{R} + \mathbf{x}(t) \quad (2.1)$$

by the second observer. Let us assume for simplicity that the axes of the two coordinate systems remain parallel (rotations are considered next). Since both observers are studying the same electronic system, the density and velocity functions must be related by:

$$n'(\mathbf{R}',t) = n(\mathbf{R},t) = n(\mathbf{R}' - \mathbf{x}(t),t)$$
$$\mathbf{u}'(\mathbf{R}',t) = \mathbf{u}(\mathbf{R},t) + \dot{\mathbf{x}}(t) = \mathbf{u}(\mathbf{R}' - \mathbf{x}(t),t) + \dot{\mathbf{x}}(t) \quad (2.2)$$

Following ideas put forth by Vignale[12], we showed in ref.[11] that in order to obtain zero XC force, we demand translational invariance i.e.

$$S[\mathbf{u}] = S[\mathbf{u}'] \quad (2.3)$$

Now, let us turn our attention to zero torque condition. Again, we refer to two observers. The first is the unprimed observer using coordinates $\mathbf{R}$ while the second is the double-primed observer. Both observers agree on the origin of coordinate systems, but not on the directions of the axes. At time $t$ a point in space labeled by the unprimed observer as $\mathbf{R}$ is labeled by the double-primed observer as:

$$\mathbf{R}'' = M(t)\mathbf{R} \quad (2.4)$$

where $M(t)$ is some instantaneous orthogonal matrix (with unit determinant) describing the mutual rotation of the axes (for convenience, we assume that $M \equiv 1$ when $t = 0$). The density and velocity fields as defined by this third observer are:

$$n''(\mathbf{R}'',t) = n(\mathbf{R},t) = n(M(t)^{-1}\mathbf{R}'',t)$$
$$\mathbf{u}''(\mathbf{R}'',t) = M(t)\mathbf{u}(\mathbf{R},t) + \dot{M}(t)\mathbf{R} \quad (2.5)$$
$$= M(t)\mathbf{u}(M(t)^{-1}\mathbf{R}'',t) + \dot{M}(t)M(t)^{-1}\mathbf{R}''$$

As for the zero XC force condition, zero total XC-torque is guaranteed when the XC action is RI,

$$S[\mathbf{u}] = S[\mathbf{u}''] \quad (2.6)$$

An action which obeys both eqs. (2.3) and (2.6) is called a GI action.

## III.  A GENERIC GI ACTION

Since the TDDFT action is unknown, we are focusing in this article on generic forms that guarantee GI, but are otherwise arbitrary. These are *plausible* forms for the action, which can serve templates for constructing approximate actions with prescribed properties. Being practical, we want to concentrate on relatively simple non-trivial generic forms. The basic idea is to first identify GI quantities accessible by TDCDFT and then writing the action *in terms of them*.

### A.  GI Action depending on the Lagrangian density

What are the simply accessible GI quantities? We follow previous works [8, 11, 13] and consider the Lagrangian coordinates, $\mathbf{R}(\mathbf{r},t)$ defined by:

$$\dot{\mathbf{R}}(\mathbf{r},t) = \mathbf{u}(\mathbf{R}(\mathbf{r},t),t) \qquad \mathbf{R}(\mathbf{r},0) = \mathbf{r} \quad (3.1)$$

$\mathbf{R}(\mathbf{r},t)$ is the position at time $t$ of a fluid element originating at a point labeled $\mathbf{r}$; in other words, $\mathbf{R}(\mathbf{r},t)$ is the *trajectory* of the fluid element $\mathbf{r}$. The coordinate $\mathbf{r}$ can be viewed as a Eularian coordinate, thus $\mathbf{R}(\mathbf{r},t)$ is the Eularian-Lagrangian transformation (ELT). Inventing memory functionals in the Lagrangian frame is easier because local memory is naturally described *within* a fluid element.

The Lagrangian density

$$N(\mathbf{r},t) = n(\mathbf{R}(\mathbf{r},t),t) \quad (3.2)$$

is GI, i.e. it is invariant with respect to both linear and rotational accelerating observers. Let us show this explicitly. Consider first accelerations. We assume both observers label the different fluid elements according to Eq. (2.1). Thus from Eq. (2.2) we see that $N(\mathbf{r},t)$ is the same by both observers, or explicitly:

$$N'(\mathbf{r},t) = n'(\mathbf{R}'(\mathbf{r},t),t) = n'(\mathbf{R}(\mathbf{r},t) + \mathbf{x}(t),t)$$
$$= n(\mathbf{R}(\mathbf{r},t),t) = N(\mathbf{r},t). \quad (3.3)$$

Note that the primed (unprimed) quantities refer to the measurements of the primed (unprimed) observer.

Now consider a rotating (double-primed) frame with fluid parcel labeling conventions given by Eq. (2.4). From Eq. (2.5), we find that $N(\mathbf{r},t)$ is RI by the following consideration:

$$N''(\mathbf{r},t) = n''(\mathbf{R}''(\mathbf{r},t),t) = n(\mathbf{R}(\mathbf{r},t),t) = N(\mathbf{r},t), \quad (3.4)$$

Eqs. (3.3) and (3.4) show that $N(\mathbf{r},t)$ is indeed GI. Thus, a simple generic form for the action functional can be immediately written down as:

$$S^{(1)}[\mathbf{u}] = s_1[N[\mathbf{u}]]. \tag{3.5}$$

Note that what this equation tells us is that the action, being a functional of the velocity field $\mathbf{u}$, must depend on it only *through* the functional dependence of $N$ on $\mathbf{u}$. This latter dependence is explicit, given by Eqs (3.1) and (3.2):

$$n_0, \mathbf{u} \to \begin{pmatrix} \mathbf{R}(\mathbf{r},t) \\ n(\mathbf{r},t) \end{pmatrix} \to N(\mathbf{r},t) \tag{3.6}$$

The functional $s_1$ in Eq. (3.5) is *any* functional of its argument. Once Eq. (3.5) is adopted, the question shifts to producing an appropriate form for $s_1$. This form is chosen so as to produce specific known physical properties of a general electronic system. Since we will generalize Eq. (3.5) in the next subsection, we will not dwell upon the form of $s_1$. We will discuss the form of the more general case.

### B. GI Action depending on the ELT metric tensor

Looking for a more general form, we now consider the Jacobian matrix of the ELT:

$$\Im_{ij} = \partial_j R_i(\mathbf{r},t) \tag{3.7}$$

Here and henceforth we use the notation $\partial_i \equiv \partial/\partial r_i$, $i = 1,2,3$. This matrix is TI, as can be straightforwardly verified[11]. However, $\Im$ is *not* RI. Indeed, the following transformation, derived from the definition of the rotation, $\mathbf{R}'' = M(t)\mathbf{R}$, must hold:

$$\Im''(\mathbf{r},t) = M(t)\Im(\mathbf{r},t) \tag{3.8}$$

While $\Im$ is not RI, its *determinant* is: since $\det \Im'' = \det M \det \Im$ and $\det M = 1$ (since it is a proper rotation matrix). One can then suggest a generic functional of the form:

$$S^{(2)}[\mathbf{u}] = s_2[\det \Im[\mathbf{u}]]. \tag{3.9}$$

Comparing with $S^{(1)}$ though, we find $S^{(2)}$ contains nothing new! This is because the function $N(\mathbf{r},t)$ is directly related to the Jacobian determinant. Indeed, the number of particles in a fluid element must be constant so $n(\mathbf{R}(\mathbf{r},t),t)d^3R = n(\mathbf{r},0)d^3r$, and thus the ratio of volume elements, which is the Jacobian is given by the ratio of densities:

$$J(\mathbf{r},t)^{-1} \equiv \left|\det[\Im(\mathbf{r},t)]\right|^{-1} = N(\mathbf{r},t)/n_0(\mathbf{r}), \tag{3.10}$$

where $n_0(\mathbf{r}) = n(\mathbf{r},0)$. Thus, the functional $s_1$ in Eq. (3.5) can also be thought of as a functional of $\det[\Im]$.

Our first attempt to introduce an action in terms of the Jacobian $\Im$ yielded nothing new (beyond Eq. (3.5)). So, let us return to Eq. (3.8), which describes how the Jacobian changes under rotations and search for another quantity which is rotationally invariant. This leads us to consider the $3 \times 3$ symmetric positive-definite ELT metric tensor:

$$g(\mathbf{r},t) = \Im(\mathbf{r},t)^T \Im(\mathbf{r},t). \tag{3.11}$$

It is immediately obvious from Eq. (3.8) and the orthogonality of $M(t)$ that $g(\mathbf{r},t) = g''(\mathbf{r},t)$. Thus $g$ is RI. And since $\Im$ is TI[11], it is also TI. We conclude that the metric tensor $g$ is GI.

One can also see that $g$ must be GI because of its geometric content: the tensor $g$ essentially tells us how to compute the distance $dS$ between two infinitesimally adjacent fluid elements, that originally started at $\mathbf{r}$ and $\mathbf{r} + d\mathbf{r}$ respectively:

$$dS^2 = \left(\mathbf{R}(\mathbf{r}+d\mathbf{r},t) - \mathbf{R}(\mathbf{r},t)\right)^2 = d\mathbf{r}^T \cdot g \cdot d\mathbf{r}. \tag{3.12}$$

Any two observers (rotated or accelerated) will agree upon such a distance between any two electron-fluid parcels. This is because the observers, while changing position of their coordinate axes origins and directions, still preserve the distance scale. Thus we conclude again that $g(\mathbf{r},t)$ *must* be GI.

The metric tensor is a natural quantity on which a generic action can defined. Thus, we consider the following class of metric-tensor actions:

$$S[n_0, \mathbf{u}] = s[n_0, g[\mathbf{u}]]. \tag{3.13}$$

Here, $n_0$ is the initial ground-state density (assuming the system starts from its ground-state). It is comforting to note that in view of Eq. (3.10) and the fact that $|\det \Im| = \sqrt{\det g}$, the generic action in Eq. (3.13) includes $S^{(1)}$ in Eq. (3.5) as a special case.

Eq. (3.13) tells us that the action depends on $\mathbf{u}$ only through the dependence of $g$ on $\mathbf{u}$. This is an explicit dependence that we designate by:

$$\mathbf{u} \to \mathbf{R} \to \Im \to g \tag{3.14}$$

Writing down Eq. (3.13) still tells us nothing about what the form of $s[n_0, g]$ is. Before we discuss this issue, we first discuss the method by which the GI potentials can be derived from the generic action in Eq. (3.13).

### IV. THE XC VECTOR POTENTIAL

In this section we set out to derive the XC potential that is obtained from the generic metric tensor action by functional derivation. It is important to realize, following van-Leeuwen's work[14], that the action and the vector potentials derived from it must be defined using a Keldysh contour. The Keldysh contour allows us to avoid causality problems inherent in any "usual" action formulated in terms of the density or current density.

The Keldysh contour is a closed loop $t(\tau)$ in time domain, parameterized by $\tau \in [0, \tau_f]$, called "pseudo-time". A closed loop, means that $t(0) = t(\tau_f)$. The use of Keldysh contours in the context of memory functionals is explained in detail in references[11, 14]. The action is formulated in terms of pseudo-time dependence $\tau$. Any physical quantity is assumed to depend on $\tau$. The potentials are then obtained as functional derivatives at the physical time dependence. By that we mean that *after* the functional derivative is taken all quantities are evaluated on the contour, replacing the variable $\tau$ by $t(\tau)$.

The potential derived from Eq. (3.13) is obtained from a chain-rule functional derivation. Defining the symmetric tensor

$$Q_{ji}(\mathbf{r}',t') \equiv 2\frac{\delta s[n_0,g]}{\delta g_{ji}(\mathbf{r}',t')}, \quad (4.1)$$

where the factor of 2 appears for later convenience. The form of the TDCDFT vector potential is obtained by considering the action change resulting from perturbing the velocity field at time $t$ and position $\mathbf{R} \equiv \mathbf{R}(\mathbf{r},t)$:

$$a_k(\mathbf{R},t(\tau)) = \frac{1}{2}\int_C \dot{t}(\tau')d\tau' \int Q_{ji}(\mathbf{r}',\tau')\frac{\delta g_{ji}(\mathbf{r}',\tau')}{\delta u_k(\mathbf{R},\tau)}d^3r', \quad (4.2)$$

(here we use the convention that repeated indices are summed over). We note that the integration on time here is performed as an integration over the Keldysh contour[14], described fully in ref[11]. All quantities that depend on time become quantities that depend on pseudo-time $\tau$. Physical quantities however are obtained after the functional differentiation by evaluating the expressions on the Keldysh contour (i.e. taking them to depend on $t(\tau)$ instead of directly on $\tau$).

The derivative appearing in Eq. (4.2) is given by:

$$\frac{\delta g_{ji}(x')}{\delta u_k(X)} = \left[\Im_{li}(x')\partial'_j + \Im_{lj}(x')\partial'_i\right]G_{lk}(x';X) \quad (4.3)$$

Where $x' \equiv (\mathbf{r}',\tau')$, $X \equiv (\mathbf{R},\tau)$ and $G_{ij}$ is derived in ref.[11], given by:

$$G_{lk}(x';X) = \left[\Im(\mathbf{r}',\tau')\Im(\mathbf{r}',\tau)^{-1}\right]_{lk} \theta(\tau'-\tau)\delta(\mathbf{R}(\mathbf{r}',\tau)-\mathbf{R}) \quad (4.4)$$

Using Eqs. (4.3) and (4.4) in (4.2), we find, integrating by parts:

$$a_k(\mathbf{R},\tau) = -\int_C \dot{t}(\tau')\int \partial_i\left[Q_{ij}(\mathbf{r}',\tau')\Im_{lj}(\mathbf{r}',\tau')\right] G_{lk}(\mathbf{r}',\tau';\mathbf{R},\tau)d^3r'd\tau' \quad (4.5)$$

Evaluating the integral on the Keldysh contour, we find the following general form vector potential:

$$\mathbf{a}(\mathbf{R}(\mathbf{r},t),t) = J(\mathbf{r},t)^{-1}\Im(\mathbf{r},t)^{-1}\mathbf{A}(\mathbf{r},t) \quad (4.6)$$

Where:

$$A_m(\mathbf{r},t) = \int_0^t \Im(\mathbf{r},t')_{ml}^T \partial_i\left[\Im(\mathbf{r},t')Q(\mathbf{r},t')\right]_{li}dt' \quad (4.7)$$

Is the "Lagrangian" vector potential.

Eqs. (3.13), (4.6) and (4.7) are the central result of this paper, resulting in a general form for a potential which yields zero force and torque. This general form should find useful application in cases where the electronic systems interact with strong fields.

Any specific XC potential derived within this framework requires specification of the functional $S[g]$ in Eq. (3.13). One source for physical information that allows construction of a proper $S[g]$ is the linear response properties of the homogeneous electron gas. This issue is studied next.

## V. XC FUNCTIONAL COMPATIBLE WITH LINEAR RESPONSE OF HOMOGENEOUS ELECTRON GAS

We would like now to compare our results with previous work on TDCDFT potentials in the linear response regime[10, 15]. For this purpose, consider Eqs. (4.6) and (4.7) developed up to first order quantities, linear in the perturbation. Consider first $\mathbf{R}(\mathbf{r},t)$, defined in Eq. (3.1). Developing it to first order, gives:

$$\mathbf{R} \to \mathbf{r} + \mathbf{R}_1 \equiv \mathbf{r} + \int_0^t \mathbf{u}(\mathbf{r},t')dt' \quad (5.1)$$

Next, consider the Jacobian matrix (Eq. (3.7)) developed to first order:

$$\Im_{ij} \to (\hat{1}+\Delta)_{ij} \equiv \delta_{ij} + \int_0^t \partial_j u_i(\mathbf{r},t')dt', \quad (5.2)$$

(where $\hat{1}$ is the $3\times 3$ unit matrix). The Jacobian determinant can be evaluated to first order using Eqs. (3.10) and (5.2):

$$J \to 1 + tr\Delta, \quad (5.3)$$

and the ELT metric tensor (Eq. (3.11)):

$$g \to \hat{1} + \Delta + \Delta^T. \quad (5.4)$$

Finally, we define for $Q_{ij}$ (Eq. (4.1))

$$Q_{ji} \to q_{ji}(\mathbf{r}) + \theta_{ji}(\mathbf{r},t) \quad (5.5)$$

where the $q(\mathbf{r})$ is a zeroth-order term of $Q$ and $\theta(\mathbf{r},t)$ is the first order term. Since $\theta$ is first order and it depends on a first order quantity, i.e. $\mathbf{u}(\mathbf{r},t)$, it can generally be written as follows:

$$\theta_{mi}(\mathbf{r},t) = \int d^3r' \int_0^t \Theta_{mi}^{kl}(\mathbf{r},\mathbf{r}'t-s)\Delta_{kl}(\mathbf{r}',s)ds. \quad (5.6)$$

Where $\Theta$ is an appropriate kernel possibly depending on $n_0$ but not on $\mathbf{u}$. Using these definitions in Eq. (4.6), we arrive at the vector potential to zero and first orders as follows:

$$\mathbf{a}_0 + \mathbf{a}_1 = \mathbf{A}_0 + \mathbf{A}_1 - (\Delta + [tr\Delta]I)\mathbf{A}_0 \quad (5.7)$$

Where the zero order term $\mathbf{A}_0$ is given by:

$$(A_0)_m(\mathbf{r},t) = t\partial_i q_{mi}(\mathbf{r}) \quad (5.8)$$

After some algebra, moving into the frequency domain, replacing $\Delta_{ji}(\mathbf{r},\omega)$ with $(i\omega)^{-1}\partial_i u_j(\mathbf{r},\omega)$ (see Eq. (5.2)) and replacing $\theta_{mi}(\mathbf{r},\omega) = \int d^3 r' \Theta_{mi}^{kl}(\mathbf{r},\mathbf{r}'\omega)\Delta_{kl}(\mathbf{r}',\omega)$ in Eq. (5.6) we find a first-order response given by the following expression:

$$[a_1]_m(\mathbf{r},\omega) = \frac{1}{\omega^2}\Big\{[\partial_m u_l + \partial_l u_m]\partial_i q_{li} + q_{li}\partial_i\partial_l u_m$$
$$+ \partial_i\left(\int \Theta_{mi}^{kl}(\mathbf{r},\mathbf{r}',\omega)\partial'_l u_k(\mathbf{r}',\omega)d^3 r'\right)\Big\} \quad (5.9)$$
$$+ \frac{\partial}{\partial\omega}\left[\frac{1}{\omega}(\partial_l u_m + \partial_k u_k \delta_{ml})\right]\partial_i q_{li}$$

Specializing to the homogeneous electron gas (HEG), it is possible to show that this form is compatible with the VK result for a HEG[10]:

$$\mathbf{a}(\mathbf{r},\omega) = \frac{1}{\omega^2}\{f_T \nabla\times[\nabla\times\mathbf{u}] - f_L \nabla[\nabla\cdot\mathbf{u}]\} \quad (5.10)$$

where $f_L(\omega,n_0)$ and $f_T(\omega,n_0)$ are the HEG response kernels. This is achieved by taking

$$\Theta_{mi}^{kl}(\mathbf{r},\mathbf{r}',\omega) = \tilde{\Theta}_{mi}^{kl}(\omega)\delta(\mathbf{r}-\mathbf{r}') \quad (5.11)$$

and having the kernels $q_{li}$ and $\tilde{\Theta}_{mi}^{kl}$ obey:

$$\tilde{\Theta}_{mi}^{kl}(\omega) + q_{li}\delta_{mk} = (\delta_{ml}\delta_{ik} - \delta_{mk}\delta_{il})f_T(\omega) - f_L(\omega)\delta_{im}\delta_{lk}$$
$$(5.12)$$

Note that for the HEG $q$ is position independent (since being a zero order quantity, it depends only on $n_0$ and $n_0$ is position independent for the HEG).

Following the general ideas of Ref.[11] for constructing an action beyond linear response, one can now choose a specific form for $s[n_0, g]$ in Eq. (3.13) as:

$$s[g] = \int g(r,t)\Theta(N(r,t'),t-t')g(r,t')d^3 rdtdt' \quad (5.13)$$

By construction, this form is fully GI (not just in the linear response limit). Furthermore, by choosing $\Theta$ in conformity with Eq. (5.12) we make this functional compatible with the linear response properties of the HEG.

## VI. SUMMARY AND DISCUSSION

Summarizing the present work, we have developed generic forms for an action which yields potentials that are consistent with the zero force and zero torque condition. The action is based on the ELT metric tensor $g$. The metric tensor is also an important quantity when the Kohn-Sham equations are presented in a Lagrangian system[16]. The metric $g$ is therefore emerging as an overall important quantity in any non-adiabatic TDCDFT scheme. The form of the action, Eq. (3.13), may also include dependence on $\dot{g}$, $\ddot{g}$ etc. In this case the functional derivatives $Q$ become differential operators with respect to time. More elaborate functionals can be built following similar ideas at the expense that additional gradients are introduced (the metric tensor already leads to potentials that use up to second gradients of the ELT $\mathbf{R}(\mathbf{r},t)$).

The present approach draws several ideas from the approach of Dobson, Bunner and Gross[13] (DBG). However, our theoretical method and our results are more general. In particular, the DBG results are consistent with only the longitudinal response of the HEG. Our method and results seem to differ significantly from another related work – namely Tokatly and Pankratov[17] (TP). The TP approach tries to construct the GI XC potentials by using the concept of a XC Lorentz force that is the divergence of a stress tensor. This latter property assures zero total force and torque, thus constituting an elegant way of enforcing these conditions. It is difficult to compare at this point the connection between these theories. It seems that the functional derivative tensor defined in Eq. (4.1) is related to the TP XC stress tensor. Further work in assessing the two theories needs to be done.

**Acknowledgements** We gratefully acknowledge the support of the German Israel Foundation.